\documentclass{iau-JDSS}
\usepackage{graphicx}

\title[How did globular clusters lose their gas?] %% give here short title %%
{How did globular clusters lose their gas?}
\author[Charbonnel, Krause {\it{et al.}}]   %% give here short author list %%
{C. Charbonnel$^{1,2}$,
 M.Krause$^{1,3,4}$, T.Decressin$^1$, G.Meynet$^1$, N.Prantzos$^5$ \and R.Diehl$^{4,3}$}

\affiliation{$^1$Geneva Observatory, University of Geneva, Versoix, Swizerland  %\break email: Corinne.Charbonnel@unige.ch \\[\affilskip]
\\[\affilskip]
$^2$IRAP, CNRS UMR 5277, Toulouse, France \\[\affilskip] $^3$ Excellence Cluster Universe, Technische Universit\"at M\"unchen, Garching, Germany \\[\affilskip]
$^4$ Max-Planck-Institut f\"ur extraterrestrische Physik, Garching, Germany \\[\affilskip] $^5$Institut d'Astrophysique de Paris, CNRS UMR7095, Paris, France }
\pubyear{2012}
\volume{Volume }  %% insert here IAU Highlights of Astronomy Volume Number
\date{?? and in revised form ??}
\jname{Special Session 1 ``Origin and Complexity of Massive Star Clusters"}
\editors{E.Vesperini \& G.Piotto, eds.}
\begin{document}

\maketitle

\begin{abstract}
We summarize the results presented in Krause et al. (2012a, K12) on the impact of supernova-driven shells and dark-remnant accretion on gas expulsion in globular cluster infancy.
\end{abstract}

\firstsection % if your document starts with a section,
\section{Introduction}
Galactic globular clusters (GCs) today appear as large aggregates of long-lived low-mass stars, with little or even no gas. Yet, they must have formed as gas-rich objects hosting also numerous massive and intermediate-mass stars. Clues on this early epoch were revealed recently with the discovery of multiple stellar generations thanks to detailed spectroscopic and deep photometric investigations (see e.g. reviews by S.Lucatello and J.Anderson, this volume). In particular, abundance data for light elements from C to Al  call for early self-enrichment of GCs by a first generation of rapidly evolving stars. Current scenarii involving either fast rotating massive stars or massive AGBs as potential polluters imply that GCs were initially much more massive and have lost most of their first generation low-mass stars (e.g. Decressin et al. 2007, 2010;  D'Ercole et al. 2008;%, 2010; 
Vesperini et al. 2010; Schaerer \& Charbonnel 2011).

\section{Driving mechanisms for gas expulsion}

Based on crude energetic arguments, gas expulsion by SNe was long thought to be the timely mechanism that could have unbound first generation stars and removed the bulk of gas together with metal-enriched SNe ejecta that are not found in second generation stars. 
In K12 we actually show that this mechanism does not generally work for typical GCs. 
Indeed, while the energy released by SNe usually exceeds the binding energy, SNe-driven shells turn out to be destroyed by Rayleigh-Taylor instability before they reach escape speed as shown on Figure 1 (left panels). Consequently the shell fragments that contain the gas remain bound to the cluster. 
This result, which is presented here for a typical protocluster of initial mass equal to $9 \times 10^6$~M$_{\odot}$ and initial half-mass radius of 3pc, holds for all but perhaps the initially least massive and most extended GCs (see K12 for details).

Instead, K12 propose that gas expulsion is launched thanks to the energy released by coherent and extremely fast accretion of interstellar gas onto dark-remnants. Due to sudden power increase in that case, the shell reaches escape speed before being affected by the Rayleigh-Taylor instability as depicted in Figure 1 (right panels). Consequently, the gas can be expelled from the cluster, and the sudden change of gravitational potential is expected to unbind a large fraction of first generation low-mass stars sitting initially in  the GC outskirts (see e.g. Decressin et al. 2010). Consequences of these results for the self-enrichment scenario will be presented in a forthcoming paper (Krause et al. 2012b, in preparation). The impact of SNe and stellar winds on their surroundings are among the current issues in understanding chemical evolution of galaxies, e.g. how interstellar gas is energized near those sources, and how ejecta are mixed into remaining gas. GCs serve as a laboratory to study this in a special, possibly extreme, environment of a smaller system, thus less complex than a galaxy as a whole.

\begin{figure*}
 \includegraphics[width=0.5\textwidth]{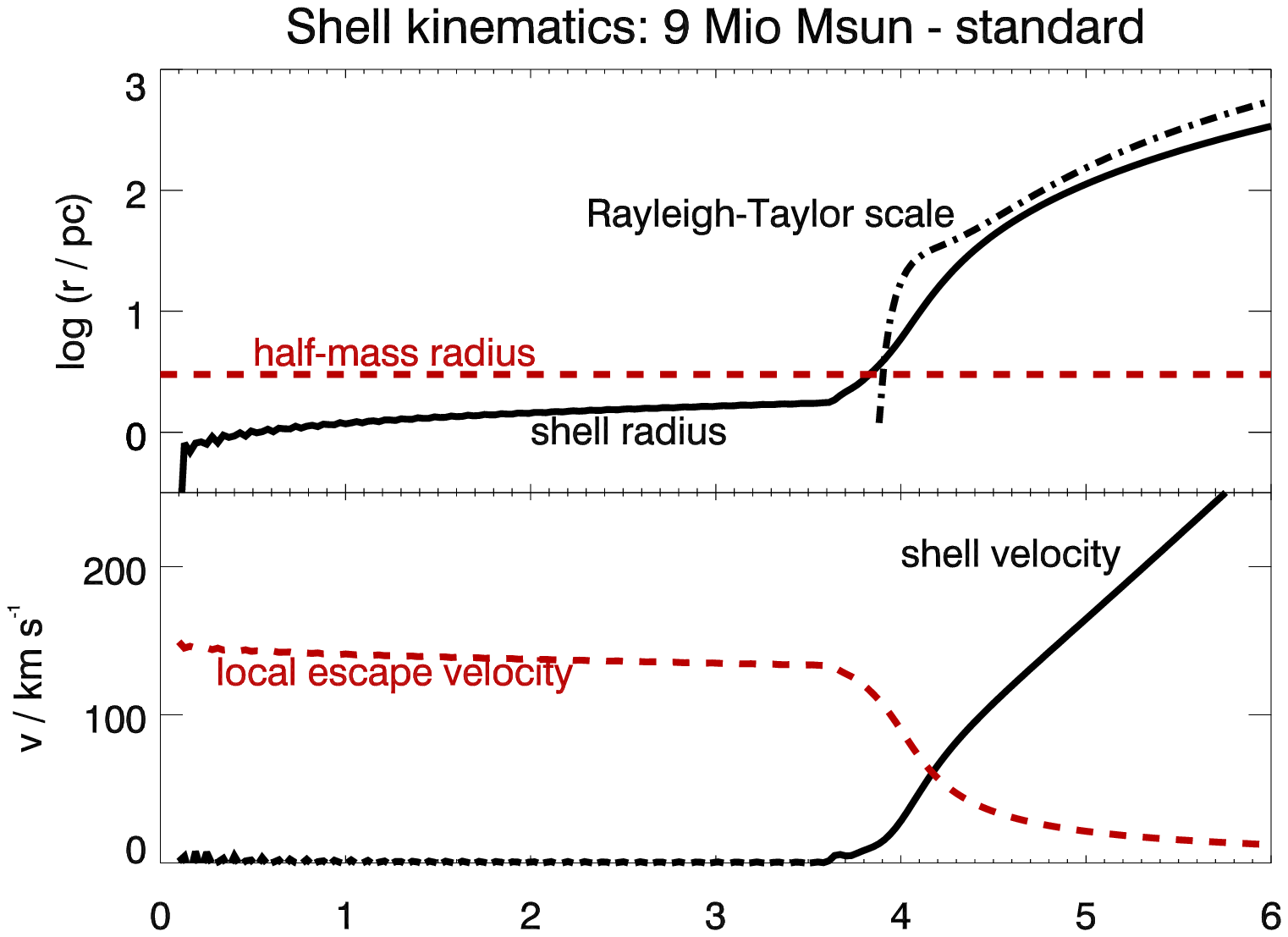}
 \includegraphics[width=0.5\textwidth]{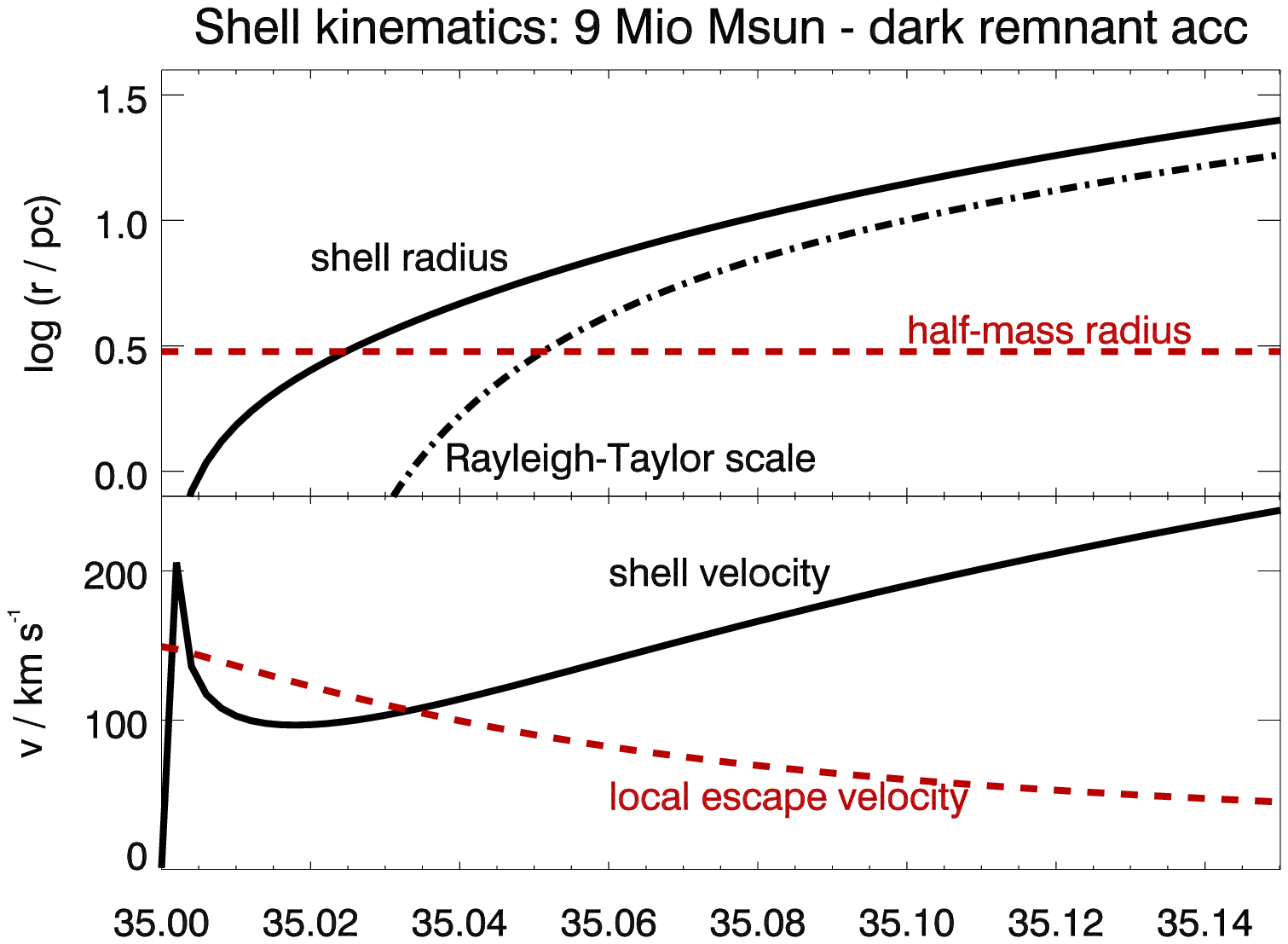}
  \caption{Superbubble kinematics within spherically symmetric thin-shell approximation. Two cases are depicted depending on the energy contributors: 
  {\it(Left)} We consider SNe explosions for all first generation stars in the mass range between 9 and 120~M$_{\odot}$, assuming that each provides 10$^{51}$erg with an efficiency parameter of 0.2. {\it(Right)} We consider sudden accretion onto all first generation dark remnants (1.5~M$_{\odot}$ neutron stars for stars between 10 and 25~M$_{\odot}$, and 3~M$_{\odot}$ black holes for more massive stars) assuming they contribute to gas energy at a rate of 20$\%$ of the Eddington luminosity.
  For both cases the upper diagrams show the bubble radius,  the Rayleigh-Taylor scale, and the GC half-mass radius (solid, dash-dotted, and dashed lines respectively), while the lower diagrams compare the shell velocity with the escape velocity at current bubble radius (solid and dashed lines respectively); note that in the left diagrams, after the Rayleigh-Taylor scale becomes larger than the shell radius, the continuous lines indicate the evolution of the shell radius and velocity  that these two quantities would have in case Rayleigh-Taylor instability would not be present.
  The timescale is initialized at the birth of first generation stars. The abscissae indicate the respective times (in Myrs) when the considered energy sources become available; in the dark remnant accretion case this is assumed here to correspond to the moment when all massive stars have died.
  \label{fig:superbublekinematics}}
\end{figure*}

\begin{acknowledgments}
We acknowledge support from the Swiss National Science Foundation, the French Society of Astronomy and Astrophysics, 
the cluster of excellence ``Origin and Structure of the Universe'', and  the ESF
EUROCORES Programme ``Origin of the Elements and Nuclear History of the Universe".
\end{acknowledgments}

\end{document}